\documentclass[twocolumn,superscriptaddress,letter]{revtex4}%
\usepackage{amssymb}
\usepackage{color}
\usepackage{graphicx}
\usepackage{dcolumn}
\usepackage{bm}
\usepackage[header,title,page,titletoc]{appendix}
\usepackage{amsmath}
\usepackage{amsfonts}%
\providecommand{\U}[1]{\protect \rule{.1in}{.1in}}

\begin{document}
\title{Ultracold Quantum Gases in Optical Lattice with Topological Defects: Its
Physics and Experimental Proposal}
\author{Xing-Hai Zhang}
\affiliation{Department of Physics, Beijing Normal University, Beijing, 100875, P. R. China}
\author{Wen-Jun Fan}
\affiliation{Department of Physics, Beijing Normal University, Beijing, 100875, P. R. China}
\author{Jin-Wei Shi}
\affiliation{Department of Physics, Beijing Normal University, Beijing, 100875, P. R. China}
\author{Su-Peng Kou}
\thanks{Corresponding author}
\email{spkou@bnu.edu.cn}
\affiliation{Department of Physics, Beijing Normal University, Beijing, 100875, P. R. China}

\begin{abstract}
In this paper we give a proposal to realize optical lattices with manipulated
dislocations and study the physics of ultracold quantum gas on a
two-dimensional (2D) optical square lattice with dislocations. In particular,
the dislocations may induce fractional topological flux on 2D Peierls optical
lattice. These results pave new approach to study the quantum many-body
systems on an optical lattice with controllable topological lattice-defects,
including the dislocations, topological fluxes.

\end{abstract}
\maketitle

\textit{Introduction}. --- Dislocations are areas where the atoms mismatch in
a perfect crystal\cite{n,cha}. There are two basic types of dislocations, the
edge dislocation and the screw dislocation. Fig.1 is the illustration of the
edge dislocations where the locus of defective points in the lattice lie along
a line. The lattice significantly distorted in the immediate vicinity of the
dislocation line. The dislocation always plays the role of impurity that
destroy the strength and ductility of metals. Thus people try to control the
dislocations to increase the quality of the metals.

Recently, people recognized that due to the interplay of defect topology and
the topology of the original states (the topological insulators, the
topological superconductors and topological orders), the dislocations will
have nontrivial quantum properties. In particular, Previous works explored the
effect of dislocations on the topological insulator, which induces zero energy
bound states~\cite{ju,ran}. Moreover, it is predicted that a Majorana fermion
zero mode may be trapped around the end of a edge dislocation in px+ipy
topological superconductor\cite{qi}. The ends of the edge dislocations obey
non-Abelian statistics, which may be applied to realize the topological
quantum computation~\cite{ki2,free,sar}. However, in condensed matter physics,
the manipulation of a dislocation in a crystal is beyond timely technology.

On the other hand, in recent years, using ultracold atoms that form
Bose-Einstein Condensates (BEC) or Fermi degenerate gases to make precise
measurements and\ simulations of quantum many-body systems has become a
rapidly-developing field\cite{Bloch's review,review}. People have successfully
observed the Mott insulator--superfluid transition in both
bosonic\cite{greiner} and fermionic\cite{Esslinger,Bloch2} degeneracy gases,
and have demonstrated how to produce\cite{Duan} and control artificial gauge
field in optical lattice \cite{ai,fl}. Without considering the harmonic trap
potential, the optical lattice can be regarded as a perfect crystal without
lattice defects such as the vacancy, the dislocation. To realize and
manipulate the topological lattice-defect in an optical lattice is an
important open issue.

In this paper we will give a proposal to realize an optical lattice with
manipulated dislocations and study the physics of ultracold quantum gas on a
two-dimensional (2D) optical square lattice with dislocations. In particular,
the dislocations may induce fractional topological flux on 2D Peierls optical
lattice (see detailed discussions below). These results pave a new way to
study the quantum many-body system on an optical lattice of ultracold quantum
gases with controllable topological lattice-defects, including the
dislocations, topological fluxes.

\begin{figure}[ptb]
\includegraphics* [width=0.48\textwidth]{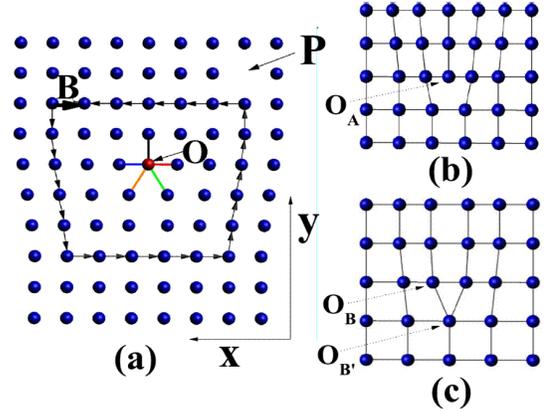}\caption{(Color online) (a)
The illustration of the dislocation, of which the end is \textrm{O} site in
its core denoted by a dotted circle. \textrm{P} site represents a vacancy that
is a non-topological defect. The Burger vector $\mathbf{B}$ is also shown -
the bold black arrow; (b) A-type dislocation of which the end is denoted by
\textrm{O}$_{A}$; (b) B-type dislocation of which the end is denoted by
\textrm{O}$_{B}$.}%
\end{figure}

\textit{The Bose-Hubbard model on 2D \textit{square} lattice with
dislocations. }--- Firstly, we focus on the edge dislocations in two
dimensions. The lattice with dislocation is shown in Fig.1 which can be
regarded as a square lattice with an line inserted. For the lattice with
dislocation along y-direction, there are five links that connect its end
denoted by \textrm{O} (See the red spot in Fig.1(a)). To describe the
topological properties of the dislocations, we define the Burger vector,
$\mathbf{B}=%
{\displaystyle \oint}
d\mathbf{u}$. For the dislocation in Fig.1(a), the corresponding Burger
vectors around the end \textrm{O} are $-\mathbf{e}_{x}$. To emphasize
distinguishing features of dislocations, we also study the non-topological
defect - a vacancy (the missing site denoted by \textrm{P} in Fig.1(a)).
Fig.1(b) and Fig.1(c) show two typical edge dislocations, A-type dislocation
and B-type dislocation. For a lattice of atoms in condensed matter physics,
B-type dislocation is always not stable. While, in optical lattices both types
of edge dislocations may exist by controlling the laser in different ways (see
below discussion).

Then we consider the one-component Bose-Hubbard (BH) model on 2D square
lattice with the dislocations shown in Fig.1(a), of which the Hamiltonian
is\cite{fisher,jaksch}
\begin{align}
H  &  =-\sum_{\langle ij\rangle}t_{ij}b_{i}^{\dagger}b_{j}+\frac{U}{2}\sum
_{i}n_{i}(n_{i}-1)\label{bh}\\
&  -\mu \sum_{i}n_{i}+h.c.,\nonumber
\end{align}
where $b_{i}^{\dag}$ and $b_{i}$ are bosonic creation and annihilation
operator on site $i$. $n_{i}=b_{i}^{\dag}b_{i}$ is the particle number
operator on site $i$. $U$ is the strength of the repulsive interaction. Except
for the lattice-site at the end of the dislocation (the red spot in Fig.1(a)),
the hopping parameters $t_{ij}$ that connect the nearest neighbor sites are
all set to be a positive number $t_{ij}=t$ ($i,j\neq$ \textrm{O}),
$\left \langle i,j\right \rangle $ represents all nearest neighboring links. For
the lattice-site at the end of the dislocation (the red spot in Fig.1), there
are five links connecting to other lattice sites. We assume the hopping
parameters on these links are $t_{ij}=t$ for the black link, $t_{ij}=t_{1}$
for the blue link, $t_{ij}=t_{2}$ for the red link, $t_{ij}=t_{3}$ for the
orange link, $t_{ij}=t_{4}$ for the green link. The local properties of the BH
model will dependent on the four distinct hopping parameters $t_{i}$
($i=1,2,3,4$). In the following, we take A-type dislocation ($t_{1}=t_{2}=t$,
$t_{3}=t_{4}=0$) and B-type dislocation ($t_{1}=t_{2}=t_{3}=t$, $t_{4}=0,$ or
$t_{1}=t_{2}=t_{4}=t$, $t_{3}=0$) as examples to study the properties of the
BH model.

This model can be solved self-consistently by the inhomogeneous mean-field
(IMF) approximation\cite{fisher,jaksch}, $b_{i}^{\dag}b_{j}\approx b_{i}%
^{\dag}\langle b_{j}\rangle+\langle b_{i}^{\dag}\rangle b_{j}-\langle
b_{i}^{\dag}\rangle \langle b_{j}\rangle,$ where $\psi_{i}=\langle b_{i}%
\rangle$ is the local superfluid (SF) order parameter (OP). Without
translation invariance, we have one local SF OP on each lattice site that
consists of the variable space of $\{ \psi_{i}\}$ on all sites. $\{ \psi
_{i}\}$ can be solved self-consistently locally from the local Hilbert space
up to $10$ one-component Bosons.

Using this IMF approach, we derive the properties of the BH model on square
lattice with dislocations. For the case of strong coupling limit, $U\gg t$,
the Bose gas turns into the Mott insulator (MI) phase; for the case of weak
coupling limit, $U\ll t$, the ground state is the SF phase with uniform OPs,
$\langle b_{i}\rangle=\psi_{i}=\psi_{0}e^{i\varphi_{0}}$ where $\psi
_{0}=\left \vert \psi_{i}\right \vert =\sqrt{n_{0}}$ and $\varphi_{0}$ is an
arbitrary real number from $0$ to $2\pi$.

Since we only have one dislocation, the global phase diagram remains the same
as that without dislocation. In the MI phase, the SF OPs vanish, $\psi_{i}=0,$
the particle number at each site remains unit for the ground state. As a
result, we find that dislocations have no significant effect in the MI phase.
On the contrary, in the SF phase, we find that near the dislocations the phase
factors of the local SF OPs on different sites are still uniform as
\textrm{Im}~($\ln \frac{\psi_{i}}{\left \vert \psi_{i}\right \vert })=\varphi
_{0}=$ constant; on the other hand, the amplitude of the local SF OPs and
local particle density change significantly near the ends the dislocation.
Near the ends of A-type dislocations \textrm{O}$_{A}$, the local particle
density decrease; while near the ends of B-type dislocations \textrm{O}$_{B}$,
the local particle density increase. See the illustrations in Fig.2(a) and
Fig.2(b). For the SF order with B-type dislocations, the local particle
density has maximum value at site \textrm{O}$_{B^{\prime}}$ rather than
\textrm{O}$_{B}$. So, the increase/decrease of local particle density near the
ends of dislocations obviously comes from the increase/decrease of the
connecting link numbers. That means the dislocations with different local
hopping parameters show different effects on the SF order locally. The changes
of local particle density $n_{i}$ near the ends of the dislocations are
strongly enhanced by the on-site interaction. Fig.2(d) shows local particle
density variation $\left \vert \Delta n/n_{0}\right \vert $ near the end of the
dislocations via $t/U$ changes where $\Delta n=n_{i}-n_{0}$. From Fig.2(d),
one can see that the on-site repulsive interaction will enhance the influence
of the lattice defects on the local particle density variation.

\begin{figure}[ptb]
\includegraphics* [width=0.52\textwidth]{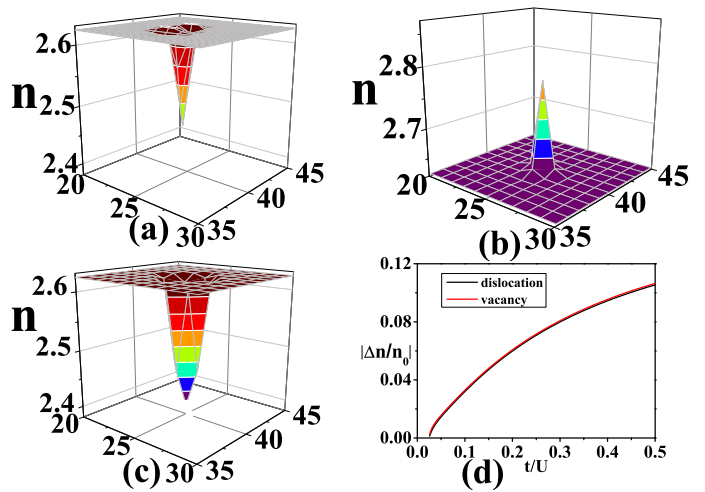}\caption{(Color online) (a)
Particle density distribution of interacting bosons near the end
(\textrm{O}$_{A}$ site) of the A-type dislocation with $t/U=0.2$ and
$\mu/U=1.5$; (b) Particle density distribution of interacting bosons near the
end (\textrm{O}$_{B}$ site) of the B-type dislocation with $t/U=0.2$ and
$\mu/U=1.5$; (c) Particle density distribution of interacting bosons near the
vacancy (P site) of the B-type dislocation with $t/U=0.2$ and $\mu/U=1.5$.
The vacancy locates at site ($x=25$, $y=40$), on which the particle density is
zero.; (d) The local particle density variation $\left \vert \Delta
n/n_{0}\right \vert $ via $t/U$ on the end of A-type dislocation (\textrm{O}%
$_{A}$) and that on the nearest site adjacency to the vacancy (missing lattice
site \textrm{P}) when $\mu/U=1.5$.}%
\end{figure}

We also studied the BH model on square lattice with a vacancy (missing
lattice-site \textrm{P} in Fig.1(a)). The results are shown in Fig.2(c). One
can see that the local SF OPs and the local particle density decrease near the
vacancies, that is similar to the case of A-type dislocations (see Fig.2(d)).
In fact, all these lattice-defects (dislocations or vacancies) lead to
\emph{small}, \emph{local} but \emph{observable} physical consequences. In the
experiments, people may detect the particle density distribution in the SF
order to observe the effect from the dislocations and that from vacancies by
In-situ Observation\cite{chin}.

\textit{The Bose-Hubbard model on 2D Peierls lattice with dislocations. }---
Next, we consider the BH on 2D Peierls lattice, of which the hopping
parameters $t_{ij}$ are not real numbers but have the uniform phase factors
along x-direction $t_{i,i+\mathbf{e}_{x}}=te^{i\phi_{x}}$ and y-direction
$t_{i,i+\mathbf{e}_{y}}=te^{i\phi_{y}}$ where $\phi_{x/y}$ is the Peierls
phase\cite{peierls1933energy} When a particle moves around a plaquette, the
extra phase factor of its wave-function is $e^{i\phi_{x}}\cdot e^{i\phi_{y}%
}\cdot e^{-i\phi_{x}}\cdot e^{-i\phi_{y}}\equiv1$.\ As a result, the non-zero
value of $\phi_{x/y}$ will lead to zero flux in each plaquette. See the
illustration in Fig.3. So, we call this type of lattice with complex hopping
parameters but no extra flux \emph{Peierls lattice}. For the free bosons, the
BEC occurs at finite momentum $\mathbf{k}=\mathbf{\phi}=(\phi_{x},$ $\phi
_{y}).$

\begin{figure}[ptb]
\includegraphics* [width=0.52\textwidth]{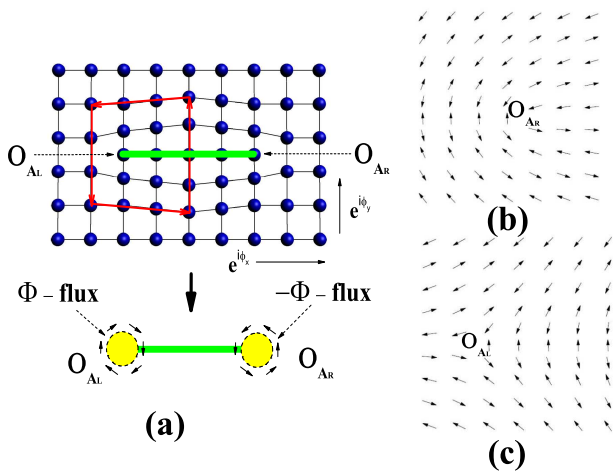}\caption{(Color online) (a)
Illustration of the induced topological flux by A-type dislocations, $\Phi
=\pm \phi_{y}$. The green line denotes the dislocation and the red lines with
arrows denote the loop path to calculate the Burger vector; (b) and (c) are
the IMF numerical results of the phase factors of local SF OPs of interacting
bosons near the ends of A-type dislocation (\textrm{O}$_{A_{L}}$ and
\textrm{O}$_{A_{R}}$) with $t/U=0.05,$ $\mu/U=1.5$ and $\mathbf{\phi}=(0,$
$\pi)$. It is obvious that the dislocation induces $\pm \pi$ topological flux
on the Peierls lattice.}%
\end{figure}

Then we study the dislocations on the Peierls lattice. An interesting property
is the \emph{induced topological flux} around the ends of the dislocations due
to the lattice mismatch. When a Boson moves around a dislocation with the
Burger vector $\mathbf{B},$ an extra phase factor will be obtained,
\begin{align}
e^{i\Phi}  &  =\left(  e^{i\phi_{x}}\cdot...\cdot e^{i\phi_{x}}\right)
_{N}\times \left(  e^{i\phi_{y}}\cdot...\cdot e^{i\phi_{y}}\right)
_{M}\nonumber \\
&  \times \left(  e^{-i\phi_{x}}\cdot...\cdot e^{-i\phi_{x}}\right)
_{N^{\prime}}\times \left(  e^{-i\phi_{y}}\cdot...\cdot e^{-i\phi_{y}}\right)
_{M^{\prime}}\nonumber \\
&  =\exp[i(%
{\displaystyle \oint}
\mathbf{\phi \cdot}d\mathbf{u)}].
\end{align}
A universal relationship between the induced topological flux and the
dislocation's Burger vector is given by
\begin{equation}
\Phi=%
{\displaystyle \oint}
d\varphi=\mathbf{\phi}\cdot \mathbf{B.} \label{relation}%
\end{equation}
Thus, the two ends of a dislocation have opposite topological fluxes, one is
$\mathbf{\phi}\cdot \mathbf{B,}$ the other is $-\mathbf{\phi}\cdot \mathbf{B}$.
The total induced topological flux of it is zero due to the cancelation effect.

Take the case in Fig.3(a) as an example. We have $N=4,$ $N^{\prime}=4,$ $M=4,$
$M^{\prime}=3$. As a result, we get a total phase factor $e^{i\Phi}%
=e^{i\phi_{y}}$. The two ends (denoted by \textrm{O}$_{A_{L}}$ and
\textrm{O}$_{A_{R}}$) of the dislocation on the Peierls lattice play the role
of a topological flux $\Phi=\pm \phi_{y}$ for all particles on the Peierls lattice.

Next, we study the properties of interacting Bosons on the Peierls lattice
with dislocations by the IMF approach. From the numerical calculations, we
found that the local SF OPs show nontrivial topological properties. For
example, for $\mathbf{\phi}=(0,$ $\pi)$, there indeed exists an $\pm \pi$ flux
around the ends of the dislocations along x-direction. From the numerical
results, the phase pattern of topological vortices near the ends
(\textrm{O}$_{A_{L}}$ and \textrm{O}$_{A_{R}}$) of the A-type dislocation
\textrm{Im}~($\ln \frac{\psi_{i}}{\left \vert \psi_{i}\right \vert })=\varphi
_{i}$ is given in Fig.3(b) and Fig.3(c). The situation is qualitatively
different from the SF order with a vacancy on the Peierls lattice. In
particular, the induced topological flux of the dislocations in the Peierls
lattice is protected by the topological properties of the dislocations. And
the fluctuations of the four distinct hopping parameters $t_{i}$ ($i=1,2,3,4$)
will never change the topological flux. As a result, the induced topological
flux from A-type dislocation is same to that from B-type.

\textit{Fermi degenerate gases on 2D square lattice with dislocations. }--- In
addition, we studied the free Fermi degenerate gases on a 2D square lattice
with dislocations. To observe the effect of the dislocation on the Fermi
liquid, we calculate the local density of states (LDOS) of the system. The
LDOS can be expressed by $N(\mathbf{r},\omega)=-\frac{2}{\pi}$\textrm{Im}%
~$[$\textrm{Tr}~$G(\mathbf{r},\mathbf{r};\omega)]$ in terms of the retarded
Green's function $G$. For the perfect lattice without dislocations, the LDOS
is uniform. When there exists a dislocation, the LDOS exhibits a quantum
interference pattern near the ends of a dislocations. Besides, the numerical
results in Fig.4 show the significant Friedel oscillations away from the ends
of a dislocations\cite{fre}. However, for the case of a vacancy, there is no
such defect induced quantum interference pattern. These predictions could also
be observed by In-situ Observation on the optical lattice with dislocations.

\begin{figure}[ptb]
\includegraphics* [width=0.5\textwidth]{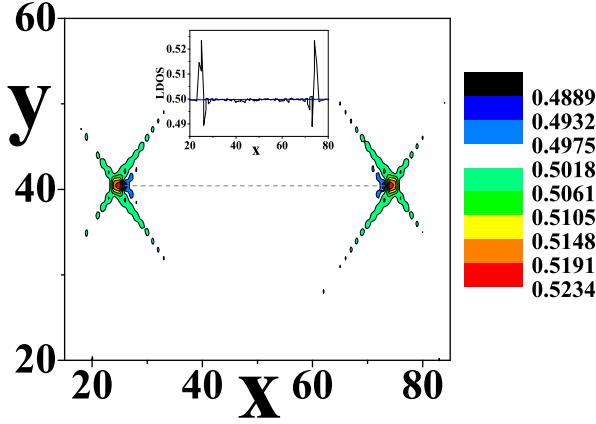}\caption{(Color online) The
local density of states of quantum interference pattern near the ends of a
dislocations for the half filling case. The system has open boundary
condition. The black dotted line denotes the dislocation. x, y denote the
lattice sites. The inset shows the oscillating local density of states along
the dislocations ($y=40$).}%
\end{figure}

However, quite different from the case of Bose gas, the induced topological
flux causes little effect on the Fermi degenerate gas.

\textit{The physical realization} --- In the followings we discuss the
physical realization of the 2D optical lattice, particularly the Peierls
lattice with dislocations.

For the first step, we give a proposal to realize an optical lattice with
dislocations by using the interference effect of the optical vortex wave. An
optical vortex is a beam of light whose phase varies in a screw thread like
manner along its axis of propagation. The optical vortex waves possess a phase
singularity which occurs at a point or a line where the physical property of
the wave becomes infinite or changes abruptly. We use the properties of
vortices to realize the dislocations \cite{Nye J F,sch,Yu N}.

An ideal optical vortex propagating in the $z$ direction may be written in the
cylindrical coordinate as $E(r,\varphi,z)=A(r,z)e^{im\varphi}e^{-ikz},$ where
$A(r,z)$ is amplitude function, $k=2\pi/\lambda$ is the wave number, and $m$
is known as the topological charge. The optical system is shown in Fig.5: a
plane wave and an optical vortex wave propagate in the $xz$ plane with
wavelength $\lambda_{1}$, linearly polarized in $y$ direction, and an optical
standing wave in the $xy$ plane with wavelength $\lambda_{2}$, linearly
polarized in $x$ direction. The optical vortex wave propagates along the $z$
axis. The plane wave in $xz$ plane is traveling in $z$ axis at an angle
$\theta$ to the optical vortex wave. The intensity of light in receiving plane
at $z_{0}$ can be written as:
\begin{equation}
I\propto \cos(k_{1}(\sin \theta x+\cos \theta z_{0}-k_{1}z_{0})+m\varphi+\phi
_{0})+\cos^{2}(k_{2}y), \label{eq:I_2D}%
\end{equation}
where $k_{1}=2\pi/\lambda_{1}$, $k_{2}=2\pi/\lambda_{2}$, $m$ is the
topological charge, $\phi_{0}$ is the phase shift from the optical path
difference between the optical vortex and the plane wave in the $xz$ plane,
$\varphi=\tan^{-1}(x/y)$. We will set the interfering plane at $z_{0}=0$ in
all our simulation. The results is shown in Fig.6. The simulation region is
$5\times5\mathrm{um}$. We have taken $\lambda_{1}=500\mathrm{nm}$,
$\lambda_{2}=600\mathrm{nm}$ in our simulations.

\begin{figure}[ptbh]
\begin{minipage}[c]{0.4\textwidth}
	  \centering
	  \includegraphics[width=1\textwidth]{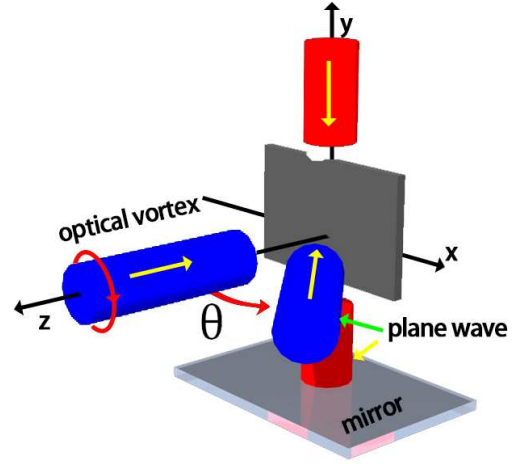}
	\end{minipage}
\caption{The experimental setup to generate an optical lattice with
dislocations: a plane wave and an optical vortex wave propagate in the $xz$
plane with wavelength $\lambda_{1},$ optical standing wave propagates in the
$xy$ plane with wavelength $\lambda_{2}$, linearly polarized in $x$
direction.}%
\end{figure}

\begin{figure}[ptbh]
\begin{minipage}[c]{0.4\textwidth}
	  \centering
	  \includegraphics[width=1\textwidth]{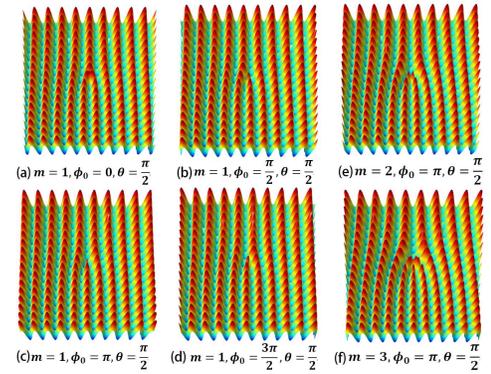}
	\end{minipage}
\caption{The intensity distributions for different dislocations with different
phase shifts in optical lattice.}%
\end{figure}

From Fig.6(a, b, c, d), we can see that the optical lattice around the
dislocation is sensitive to the phase shift between the optical vortex and the
plane wave. So we need to adjust the phase shift properly to obtain the
designed dislocation optical lattice, which can be realized by tunable phase
plate. The fringe spacing would increase due to the decrease of the angle
$\theta$. So an A-type dislocation (Fig.6(a)) can evolute into a B-type
(Fig.6(d)) one \emph{smoothly} by tuning the angle $\theta$. This procedure
also indicates the topology-equivalence between A-type and B-type
dislocations. In addition, we can also generate a dislocation with higher
Burger vector if the topological charge $|m|>1$, as shown in Fig.6(e, f).

For the second step we show how to realize the Peierls optical lattice.
Recently, in experiments, tunable Peierls phases in a 1D \textquotedblleft
Zeeman lattice\textquotedblright \ have been realized using a combination of
radio-frequency and Raman coupling \cite{jimenezGarcia2012peierls}. Now, a hot
topic is to realize the spatially-dependent Peierls phases that leads to an
effective (staggered) magnetic flux through one unit cell of the optical
lattice. Thus, to engineer uniform complex hopping amplitudes with Peierls
phases along y-direction\cite{peierls1933energy}, $t_{i,i+\mathbf{e}_{y}%
}=te^{i\phi}$, the Raman-assisted hoppings are used by a pair of far-detuned
running-wave laser beams with different frequencies, $\omega_{0}$ and
$\omega_{0}+\Delta$%
\cite{Eckardt2005localization,Lignier2007,Chen2011controlling,kolovsky2011creating}%
. A linear potential is added along y-direction, of which the energy
difference between two nearest neighbor sites is $\Delta.$ So the
Raman-assisted hoppings are restored along y-direction with the energy
resonance $\hbar \omega=\Delta$. However, along x-direction, there is no such
Raman-assisted hoppings. As a result, we can get uniform complex hopping
amplitudes with Peierls phases along y-direction, $t_{i,i+\mathbf{e}_{x}}=t,$
$t_{i,i+\mathbf{e}_{y}}=te^{i\phi_{y}}$. By similar approach and tuning the
directions of the two Raman laser beams, we can get different uniform complex
hopping amplitudes with Peierls phases along both x-direction and y-direction.

\textit{Conclusion --- }In this paper we studied the physics of ultracold
quantum gases on a two-dimensional (2D) optical square lattice with
dislocations. We found that for the Boson's SF order in conventional 2D square
lattice, the dislocations will slightly change the local particle density and
lead to small, local but observable physical consequences. In particular, the
dislocations may induce fractional topological flux on 2D Peierls optical
lattice from the interplay between the Peierls phases on hopping amplitudes
and the nonlocal properties of the dislocations. In this case, the topological
optical vortex eventually results the topological vortex of SF of Bosons in
optical lattice. In the future, this realization may be applied to realized
nontrivial topology in topological ordered states.

\begin{center}
{\textbf{* * *}}
\end{center}

This work is supported by National Basic Research Program of China (973
Program) under the grant No. 2011CB921803, 2012CB921704 and NSFC Grant No.
11174035 and 11374037.

\end{document}